\documentclass[leqno, notitlepage]{amsart}
\usepackage{amscd}
\usepackage{amssymb}
\usepackage[all, poly]{xy}

\DeclareMathOperator{\re}{re}
\DeclareMathOperator{\Hom}{Hom}

\DeclareMathOperator{\Irr}{Irr}

\DeclareMathOperator{\Ind}{ind}

\DeclareMathOperator{\ad}{ad}

\DeclareMathOperator{\al}{\alpha}
\DeclareMathOperator{\e}{\epsilon}
\DeclareMathOperator{\MLZ}{\mathcal L_{\mathbb Z}}

\DeclareMathOperator{\E}{\bf E}
\DeclareMathOperator{\C}{\mathbb C}
\DeclareMathOperator{\R}{\mathbb R}
\DeclareMathOperator{\Q}{\mathbb Q}
\DeclareMathOperator{\Z}{\mathbb Z}

\DeclareMathOperator{\ML}{\mathcal L}
\DeclareMathOperator{\pt}{pt}
\DeclareMathOperator{\Card}{Card}

\DeclareMathOperator{\tha}{\theta}

\DeclareMathOperator{\Om}{\Omega}
\DeclareMathOperator{\OOm}{\overline{\Omega}}
\DeclareMathOperator{\Th}{\Theta}

\DeclareMathOperator{\Cl}{Cl}
\DeclareMathOperator{\cl}{cl}
\DeclareMathOperator{\sgn}{sgn}

\DeclareMathOperator{\height}{ht}

\DeclareMathOperator{\n}{\mathfrak n}

\newtheorem*{definition}{Definition}

\swapnumbers
\theoremstyle{plain}

\newtheorem*{conjecture}{Conjecture}
\newtheorem*{lemma}{Lemma}
\newtheorem*{proposition}{Proposition}

\numberwithin{equation}{subsubsection}

\numberwithin{enumi}{subsubsection}

\begin{document}

\title[Quiver varieties, BPS states, semicanonical basis]
{Quiver varieties, affine Lie algebras, algebras of BPS states, 
and semicanonical basis}

\author{ Igor Frenkel }
\address{Department of Mathematics, Yale University,
10 Hillhouse Ave, New Haven, CT 06520}

\email{frenkel@math.yale.edu}

\author{ Anton Malkin }
\address{Department of Mathematics, MIT,
77 Massachusetts Avenue, Cambridge, MA 02139}

\email{malkin@math.mit.edu}

\author{ Maxim Vybornov }
\address{Department of Mathematics, 
University of Massachusetts, 
Lederle Graduate Research Tower, 
Amherst, MA 01003}

\email{vybornov@math.umass.edu}

%\date{May 13, 2002}

\begin{abstract} 
We suggest a (conjectural) construction of 
a basis in the plus part of the affine
Lie algebra of type ADE indexed by irreducible components of
certain quiver varieties. 
This construction is closely related to 
a string-theoretic construction of a Lie algebra of BPS states.
We then study the new combinatorial questions about the 
(classical) root systems naturally arising from our 
constructions and Lusztig's semicanonical basis. 
\end{abstract}

\maketitle

\section{Introduction}

\subsection{} Since the appearance of the first manifestation of
the connection between theory
of representations of quivers and the structure theory
of Lie algebras \cite{Gabriel1972}, 
several authors discovered constructions of Lie algebras
arising from the quiver theory. 
C. M. Ringel constructed the plus part of the simple Lie algebra
of type ADE, \cite{Ringel1990b}, and 
a (related) construction of the plus part of any Kac-Moody 
Lie algebra is implicit
in Lusztig's construction of (quantized) enveloping algebras
of Kac-Moody Lie algebras in terms of functions on a class of 
remarkable affine varieties $\Lambda_V$, 
\cite{Lusztig1991a, Lusztig1992}. 

Inspired by Ringel and Lusztig, 
the present authors came up with a construction
of the plus part $\widetilde{\n}_{+}$ of the affine Lie algebra of type ADE
in terms of indecomposable representations of quivers \cite{FMV00}.

A similar construction is suggested by string theorists 
\cite{FM00}, who use
the \emph{stable} representations of (double) quivers rather than
indecomposable representations of oriented quivers (as do Ringel and
the present authors). The concept of stability has been very useful in
representation theory:  H. Nakajima discovered that modules over 
Kac-Moody Lie algebras
may be described using functions on the varieties $\Lambda^{s}_V/G_{V}$
where $\Lambda^{s}_V$ are the \emph{stable} points of Lusztig's 
varieties $\Lambda_V$
\cite{Nakajima1994, Nakajima1998}. 

\subsection{} One of the main goals of this paper is to study the 
relationship between the Lusztig's construction of $\widetilde{\n}_{+}$,
the construction by the present authors \cite{FMV00}, and the ``stable''
construction suggested by physicists. 
Using the notion of \emph{semistable diagonal} in $\Lambda_V$
suggested in \cite{HM98},
we make a conjecture 
(Conjecture \ref{conjecture})
directly relating the Lusztig's construction and the ``stable''
construction. Moreover, we (conjecturally) obtain a basis in 
$\widetilde{\n}_{+}$
parameterized by irreducible components of algebraic varieties.
One can look at this conjectural basis as ``semicanonical''
basis for $\widetilde{\n}_{+}$.

\subsection{} 
In section \ref{semicanonical}
we use our methods to obtain some simple results 
describing new aspects (arising from quiver constructions) 
of combinatorics of the root systems and Weyl groups, 
and ask many more questions 
than we can answer at the moment. 

One of the results of the Lusztig construction
of the (quantized) universal enveloping algebras in terms of functions 
on the varieties $\Lambda_V$ is the appearance of the 
\emph{semicanonical} basis in the non-quantized enveloping algebras
of simply laced Kac-Moody Lie algebras \cite{Lusztig1999}.
Semicanonical basis is indexed by the irreducible components of $\Lambda_V$.
It was expected that this basis would
coincide with the specialization of the Lusztig's canonical basis to $q=1$,
but a counterexample was found by M. Kashiwara and Y. Saito
\cite{KS1997}. The relationship between these two bases is quite mysterious 
at the moment, even though they share their combinatorial properties.
 
Let $\n_{+}\subset U(\n_{+})$ be the plus part of 
a simple Lie algebra of type ADE, and 
$U(\n_{+})$ be its universal enveloping algebra. For every
positive root $\al\in R_{+}$ we have a canonically (up to a sign)
defined element $E^{*}_{\al}\in \n_{\al}\subset U(\n_{+})$, where
$\n_{\al}$ is the one dimensional root subspace of $\n_{+}$ 
corresponding to $\al$. 
We study the decomposition of $E^{*}_{\al}$ with respect
to the semicanonical basis. 
More precisely,
let us consider $\Lambda_V$, 
$\dim V=\al$. Then the irreducible components of $\Lambda_V$
may be indexed by decompositions of $\al$ into the sum
of positive roots. Let $\al=\al_1+\al_2+\dots+\al_l$, $\al_i\in R_{+}$
be such a decomposition and let $e_{\al_1+\al_2+\dots +\al_l}$ be the
element of the semicanonical basis corresponding to this
decomposition (and the corresponding irreducible component of $\Lambda_V$).
Then
$$
E^{*}_{\al}=\sum_{\al=\al_1+\al_2+\dots +\al_l}
c_{\al_1+\al_2+\dots +\al_l}e_{\al_1+\al_2+\dots +\al_l} ,
$$
where $c_{\al_1+\al_2+\dots +\al_l}\in \Z$ is the coefficient of
$e_{\al_1+\al_2+\dots +\al_l}$. In other words, with the help of
the semicanonical basis we assign an integer $c_{\al_1+\al_2+\dots +\al_l}$
to every decomposition $\al=\al_1+\al_2+\dots +\al_l$ of the root
$\al$ into a sum of positive roots. It would be very interesting
give a purely combinatorial description of these numbers 
without appealing to the semicanonical basis. We manage to
obtain such a description in the $A_n$ case in terms of the sign character
of the Weyl group $W=S_{n+1}$.

\subsection{Acknowledgment} We are grateful to Naihuan Jing
who presented us with the opportunity to give this talk
at the CBMS conference he organized at NCSU in June 2001. 
We tried to preserve a bit
of the informal style of a talk in this paper.
We are grateful to B. Fiol, J. Humphreys, G. Lusztig, M. Mari\~ no, 
and C. Ringel
for useful conversations. 
M.V. is grateful to MSRI and Max-Planck-Institut f\" ur Mathematik, 
where parts of this work were done, for their hospitality.
The research of I. F. was supported in part by NSF.
The research of A. M. and M. V. was supported by NSF Postdoctoral 
Research Fellowships. 

\section{Preliminaries}

We always assume that our ground field is the field
of complex numbers $\C$. Our notation and conventions are mostly 
lifted from \cite{Lusztig1991a, Lusztig1999, FMV00}.

\subsection{Quivers}
\subsubsection{} 
To a graph $Q$, with no edges joining a vertex with itself, 
we associate a pair of sets: 
$I$ (vertices), and $H$ (oriented edges), and
two maps from $H$ to $I$:
\begin{enumerate}
\item a map $H\to I$ denoted $h\to h'$ (initial vertex),
\item a map $H\to I$ denoted $h\to h''$ (terminal vertex),
\item  an involution  $h\to \bar h$ on $H$ which maps
an oriented edge to the same edge with the opposite orientation.
\end{enumerate}

An \emph{orientation} of $(I,H)$ is a choice of a subset $\Om\subset H$
such that $\Om\cup \OOm =H$ and 
$\Om\cap \OOm =\emptyset$. 
Abusing terminology we will call
both $(I,H)$ and $(I,\Om)$ \emph{quivers}. 

\subsubsection{}\label{reps}
Let $\mathcal V$ be the category of $I$-graded vector spaces
$V=\oplus_{i\in I}V_i$.
We define:
$$
\begin{aligned}
\E_V=& \bigoplus_{h\in H}\Hom(V_{h'},V_{h''}),\\
\E_{V,\Om}=& \bigoplus_{h\in \Om}\Hom(V_{h'},V_{h''}).\\
\end{aligned}
$$
For an element $x\in\E_V$ we denote by $x_h$, $h\in H$ its component
in $\Hom(V_{h'},V_{h''})$.
A pair $(V,x)$, $x\in\E_V$ (resp. $x\in\E_{V,\Om}$) 
is called a representation of $(I,H)$ (resp. $(I,\Om)$).
We will also sometimes call $(V,x)$ a module over $(I,H)$,
and moreover denote such a module by $V$ if it is clear 
what $x$ we consider.

Let $\Z[I]$ be the free abelian group generated by the set $I$.
The \emph{dimension} of $(V,x)$ with $x\in\E_V$ or $x\in\E_{V,\Om}$
is an element of $\Z[I]$ defined as follows:
$$
\dim (V,x)=\dim V=\sum_{i\in I}\dim_{\C} (V_i) i\in\Z[I].
$$  

The algebraic group $G_V=\prod_{i\in I}GL(V_i)$ acts on $\E_V$
in a natural way. 

\subsubsection{}\label{indecomp} We say that $x\in\E_{V,\Om}$ is
\emph{indecomposable} if $(V,x)$ is indecomposable as
a representation of $(I,\Om)$. The subset of indecomposable
elements in $\E_{V,\Om}$ is denoted by $\E^{\Ind}_{V,\Om}$.

\subsubsection{}\label{nilpotent} 
We say that $x\in\E_{V}$ is \emph{nilpotent} if there exists
an $N\geq 2$ such that for any sequence $h_1,h_2,\dots,h_N\in H$, 
such that $h_1'=h_2'', h_2'=h_3'',\dots, h_{N-1}'=h_N''$, the 
composition $x_{h_1}x_{h_2}\dots x_{h_N}=0$. 
The subset of nilpotent
elements in $\E_{V}$ is denoted by $\E^{nil}_{V}$.

\subsubsection{}\label{moment}  
Following \cite{Lusztig1991a} we consider
the moment map attached to the $G_V$-action on $\E_V$.
The $i$-component $m_i$ of this map is given by
$$
m_i(x)=\sum_{h\in H:h''=i} \epsilon(h)x_{h}x_{\bar h} ,
$$ 
where $\epsilon: H\to \C^*$ is a function such that 
$\epsilon(h)+\epsilon(\bar h)=0$ for all $h\in H$. 
Following Lusztig we introduce the subvariety
$\Lambda_V$ of $\E_V$ as follows

\begin{definition} $\Lambda_V$ is the closed subvariety of 
all nilpotent elements 
$x\in\E_V$ such that $m_i(x)=0$ for all $i\in I$. 
\end{definition}

\subsection{Convolution product}

\subsubsection{}  
Let 
$$
\ML(V)= M_{G_V}(\Lambda_V) 
$$
be the set of constructible
$\C$-valued functions on $\Lambda_V$ which are constant
on $G_V$-orbits in $\Lambda_V$. 
Let $\ML_{\Q}(V)$ (resp. $\MLZ(V)$) 
be the set of all $f\in\ML(V)$ with rational (resp. integral) values. 

\subsubsection{}\label{convolution}
Let $V, V', V''\in\mathcal V$
be such that $\dim V'+\dim V''=\dim V$. Let $f'\in\ML(V')$,
$f''\in\ML(V'')$. We lift the convolution
construction from \cite[12.10]{Lusztig1991a}.
Consider the diagram
$$
\begin{CD}
\Lambda_{V'}\times\Lambda_{V''}@<{p_1}<<{\bf F'} @>{p_2}>>{\bf F''}
@>{p_3}>>\Lambda_V ,
\end{CD}
$$
where $\bf F''$ is the variety of all pairs $(x,W)$ where $x\in\Lambda_V$ 
and $W\subset V$ is an $x$-stable subspace, $\dim W=\dim V''$.

$\bf F'$ is the variety of all quadruples $(x,W,R'',R')$ where
$(x,W)\in\bf F''$, and $R', R''$ are $\mathcal V$-isomorphisms
$R': V'\simeq V/W$, $R'':V''\simeq W$.

We have $p_1(x,W,R'',R')=(x',x'')$ where 
$x_hR'_{h'}=R'_{h''}x'_h$, and $x_hR''_{h'}=R''_{h''}x''_h$ for all $h\in H$. 

We have $p_2(x,W,R'',R')=(x,W)$, $p_3(x,W)=x$.

Let $f_1\in\ML(\Lambda_{V'}\times\Lambda_{V''})$ be given by
$f_1(x',x'')=f'(x')f''(x'')$. Then there is a unique function 
$f_3\in\ML(\bf F'')$ such that $p_1^*f_1=p_2^*f_3$. Then by definition
$$
f'*f''=(p_3)_!(f_3).
$$

There exists an analogous construction for the oriented quiver 
\cite[10.19]{Lusztig1991a}. We will denote the corresponding convolution 
product by $*_{\Om}$ for an orientation $\Om$.

\subsubsection{}\label{defl0}
If $\dim V=i\in I$, then $\Lambda_V$ is a point, and we denote
by $E_i(\Lambda_V)\equiv 1$ the function which is identically 
$1$ on this point.  
Let $\ML_0$ be the associative algebra with $*$-product   
generated by $\{E_i\}_{i\in I}$. (The associative algebra 
of functions on $\E_{V,\Om}$ 
with $*_{\Om}$-product   
generated by $\{E_i\}_{i\in I}$ will be denoted by $\ML_{0,\Om}$.)
One can consider $\ML_0$ as a $\mathbb Z_+[I]$-graded Lie algebra over
$\mathbb Q$, with the following Lie bracket:
$$
[f,g]=f*g-g*f.
$$
We denote by $\n^*$ the Lie subalgebra of $\ML_0$ generated by
$\{E_i\}_{i\in I}$.

\subsection{Geometric realization of the enveloping algebra}
\label{georealization}
\subsubsection{}
Let $(I,H)$ be a quiver. For $i,j\in I$ define 
$i\cdot j=-\Card\{h\in H\ |\ h'=i, h''=j\}$, if $i\neq j$ and
$i\cdot j=2$ if $i=j$.
Let $U^+$ be the $\C$-algebra defined by generators 
$e_i$, $i\in I$ and the Serre relations:
$$
\sum_{\substack{p,q\in\Z_{\geq 0} \\ p+q=-i\cdot j+1}}(-1)^{p}
\frac{e_i^p}{p!}e_j\frac{e_i^q}{q!}=0
$$
for any $i\neq j$ in $I$. $U^+=U(\n)$ is the 
enveloping algebra of the Lie algebra
$\n$ defined by generators 
$e_i$, $i\in I$ and the Serre relations:
$$
(\ad e_i)^{-i\cdot j+1}(e_j)=0
$$
for any $i\neq j$ in $I$.

Let $(V,x)$ be a representation of $(I,H)$. 
Let $U^+_{\dim V}$ denote the subspace of $U^+$
generated by $e_{i_1}e_{i_2}\dots e_{i_n}$ for sequences
$i_1,i_2,\dots,i_n$ in which $i$ appears $\dim V_i$ times for any $i\in I$.

Let $U^+_{\Z}$ be the subring of $U^+$ generated
by the elements $e_i^p/p!$ for all $i\in I$, $p\in\Z_{\geq 0}$.
Then $U^+_{\Z, \dim V}=U^+_{\Z}\cap U^+_{\dim V}$, see 
\cite[1.1]{Lusztig1999}.

\subsubsection{}\label{psi}
In \cite[12.12]{Lusztig1991a} Lusztig defines a 
$\mathbb C$-linear map $\psi_V:U^+_{\dim V}\to \ML(V)$, such that
\begin{enumerate}
\item $\psi=\bigoplus_{\dim V}\psi_V : U^+\simeq\ML_0$ is an
isomorphism of algebras,
\item $\psi(e_i)=E_i$, for any $i\in I$,
\item $\psi_V(U^+_{\Z, \dim V})\subset\MLZ(V)$.
\end{enumerate}  
The isomorphism $\psi$ also restricts
to an isomorphism of Lie algebras $\psi:\n\to \n^{*}$
such that $\psi(e_i)=E_i$ for any $i\in I$.

In \cite{Lusztig1999} Lusztig defines a 
$\mathbb Q$ version of the map  $\psi_V:U^+_{\dim V}\to \ML(V)$, 
which restricts to a map 
$\psi_V: U^+_{\mathbb Z,\dim V}\to \ML_{\mathbb Z}(V)$.

\subsection{Lie algebra based on the Euler cocycle: classical ADE case}
\label{LieEulerfinite}
Let the graph underlying the quiver $(I,H)$ be the
Dynkin diagram of type ADE. In this case we can identify 
$\Z[I]$ with the root lattice of type ADE. The elements $i\in I$
are considered simple roots, and we have the root system 
$R\subset \Z[I]$ and its positive part $R_{+}\subset\Z_{\geq 0}[I]$.

Let us fix a cocylcle $\e: \Z[I] \times \Z[I]
\rightarrow \Z/{2\Z}$.
In the simply laced Dynkin diagram case, the Lie algebra
$\n=\n^{*}$ defined above is isomorphic to the Lie algebra $n^{\e}$ 
spanned by the elements $\tilde e_{\alpha}$, 
$\alpha\in R_{+}$ with the bracket defined by $\e$ as in 
\cite[4.2.1.a]{FMV00}.

Now let us define the functions $E^{*}_{\alpha}\in \ML(V)$, $\alpha\in R_{+}$,
$\dim V=\alpha$ as follows:
\begin{equation}
E^{*}_{\alpha}=\psi(\tilde e_{\alpha}).
\end{equation}

Notice that if $\alpha=i\in I$, then $\Lambda_V$ is a point and 
$E^{*}_i=E_i$. 

\subsection{Lie algebra based on the Euler cocycle: affine ADE case}
\label{LieEuleraffine}
Let the graph underlying the quiver $(I,H)$ be the extended
Dynkin diagram of type ADE. In this case we can identify 
$\Z[I]$ with the affine root lattice of type ADE. The elements $i\in I$
are considered simple roots, and we have the affine root system 
$R\subset \Z[I]$ and its positive part $R_{+}\subset\Z_{\geq 0}[I]$.

Let us choose an extending vertex $p\in I$,
and let $I'=I-\{p\}$.

Again, let us fix a cocylcle $\e: \mathbb{Z} [I] \times \mathbb{Z} [I]
\rightarrow \mathbb{Z}/2\Z$. Let $\n_{\alpha}$ be the root space
of the algebra $\n$ corresponding to $\alpha\in R_{+}$.
Let $\delta$ be the indivisible imaginary root.
The Lie algebra $\n$ is isomorphic to the Lie algebra $\n^{\e}$ 
spanned by the elements $\tilde e_{\alpha}\in \n^{\e}_{\alpha}$, 
$\alpha\in R^{\re}_{+}$, and $\alpha_k(n)\in \n^{\e}_{n\delta}$, 
$n\geq 1$, where we identify 
$\n^{\e}_{n\delta}=\C[I']=\text{the vector space spanned by }I'$,
and $\alpha_k=k\in I'\subset \mathbb Z[I']$ are simple roots.
The bracket is defined by $\e$ as in 
\cite[5.2.1.a]{FMV00}. In particular,

\begin{equation}\label{EpsilonCommutatorAffine}
\begin{split}
[\Tilde{e}_\alpha, \Tilde{e}_\beta] &= 
\begin{cases}
\epsilon (\alpha , \beta ) 
\Tilde{e}_{\alpha + \beta} &
\text{ if } \alpha + \beta \in R^{\re}_+ , \\ 
\epsilon (\alpha , \beta ) 
\alpha (n) &
\text{ if } \alpha + \beta = n \delta , \\ 
0 & 
\text{ if } \alpha + \beta \notin R_+ . 
\end{cases}
\end{split}\end{equation}

Now let us define the functions $E^{*}_{\alpha}\in \ML(V)$, 
$\dim V=\alpha\in R^{\re}_{+}$ as follows:
\begin{equation}
E^{*}_{\alpha}=\psi(\tilde e_{\alpha}) ,
\end{equation}
and the functions $E^{*}_k(n)\in \ML(V)$, $\dim V=n\delta$ as follows:
\begin{equation}
E^{*}_k(n)=\psi(\alpha_k(n)) .
\end{equation}
Again, if $\alpha=i\in I$, then $\Lambda_V$ is a point and 
$E^{*}_i=E_i$. 

\section{Stability and simple Lie algebras of type ADE}\label{stablesection}

\subsection{Stability after King and Rudakov}

\subsubsection{Stability after King} 
Let us consider the abelian 
category of nilpotent representations of a quiver $(I,H)$ and its Grothendieck
group $K_0$. A \emph{character}
on $K_0$ is an additive function $\tha: K_0\to\R$.  

\begin{definition}{\label{stabilityk}}\cite{King1994}
A point $x\in\E_V$ is called $\tha$-stable (resp. 
$\tha$-semistable) 
if $\tha(V)=0$, and  
for any $x$-stable nonzero proper subspace 
$V'\subset V$ we have $\tha(V')>0$ (resp. $\tha(V')\geq 0$).
\end{definition}

\begin{definition} Two $\tha$-semistable points $x',x''\in\E_{V}$ 
are \emph{$S$-equivalent} (notation: $x'\overset{S}\sim x''$) 
if the orbit closures 
$\overline{G_V\cdot x'}\cap\overline{G_V\cdot x''}$ intersect
in the set of $\tha$-semistable points $\E_V^{ss}$.
\end{definition}

\subsubsection{Stability after Rudakov} 
One can also define stable points using 
a ``slope'' stability condition: $\mu=c/r$ where $c$ and 
$r$ are additive functions $K_0\to\R$ and $r(V)>0$ for any 
$V\in\mathcal V$, see \cite[3]{Rudakov1997}.

\begin{definition}{\label{stabilityr}}\cite{Rudakov1997}
A point $x\in\E_V$ is called \emph{$\mu$-stable} (resp. 
\emph{$\mu$-semistable}) if for any $x$-stable 
nonzero proper subspace $V'\subset V$ we have
$\mu(V')<\mu(V)$ (resp. $\mu(V')\leq \mu(V)$).
\end{definition}

If we fix such slope stability condition $\mu$ and $V\in\mathcal V$ 
we can define a character $\tha_{\mu}: K_0\to\R$ as follows:
\begin{equation}
\tha_{\mu}(V')=-c(V')+\frac{c(V)}{r(V)}r(V').
\end{equation}
According to \cite[Proposition 3.4]{Rudakov1997} a point $x\in\E_V$
is $\mu$-stable if and only if it is $\tha_{\mu}$-stable.
 
\subsubsection{}\label{king}
Let us fix an orientation $\Om\subset H$ until the end of this section.
Following \cite{King1994}
we construct a character $\tha=\Th_{V,\Om}$
associated to $\Om$.
Let $V,V'\in \mathcal V$.
We define
$$
\Th_{V,\Om}(V')=\sum_{h\in\Om}(\dim V_{h'}\dim V'_{h''}-
\dim V'_{h'}\dim V_{h''}).
$$

\subsection{Stability Lemma}
In this section we will consider quivers of finite ADE type, i.e. 
the underlying non-oriented graph $Q$ is a Dynkin graph
of (finite) ADE type. A fact similar to the lemma below 
was independently conjectured by
M. Reineke \cite[Conjecture 7.1]{Reineke}

\begin{lemma}\label{stablemma}
Let $x\in\Lambda_V$, $\dim V=\alpha\in R_{+}$. 
Then $x$ is $\Th_{V,\Om}$-stable if and only if
$x\in\E^{\Ind}_{V,\Om}$.
\end{lemma}

\begin{proof} Since in the Dynkin quiver case every root is a Schur root
\cite{King1994}, a generic point in $\E_{V,\Om}$ is $\Th_{V,\Om}$-stable
according to \cite{King1994, Schofield1992}.
Since $\E^{\Ind}_{V,\Om}$ is an open dense $G_V$-orbit in 
$\E_{V,\Om}$, any $x\in\E^{\Ind}_{V,\Om}$ is $\Th_{V,\Om}$-stable.

Let us prove the ``only if'' part. If $x\in\Lambda_V-\E^{\Ind}_{V,\Om}$, 
then
we have $x=y+z$, where $y\in\E_{V,\Om}$, $z\in\E_{V,\OOm}$.
Following \cite[14]{Lusztig1991a} we see that there exists
a decomposition:
$$
(V,y)=\oplus_{p=1}^{\nu}(V^p,y)
$$
such that $\Hom(V^p,V^{p'})=0$ whenever $p'<p$.
There is a direct sum decomposition:
$$
\E_{V,\Om}=\bigoplus_{1\leq p,p'\leq \nu}\E^{p,p'}_{V,\Om}
\qquad {\text{where}}\qquad  
\E^{p,p'}_{V,\Om}=\Hom_{h\in\Om}(V_{h'}^p, V_{h''}^{p'}).
$$

Now we need the following claim: there exists 
$y'\in\bigoplus_{p> p'}\E^{p,p'}_{V,\Om}$ 
such that $y+y'=:x'\in\E^{\Ind}_{V,\Om}$.
It is not very hard to prove this claim using the
methods of \cite[4]{Lusztig1990}.

Following \cite[14]{Lusztig1991a} we denote 
$V^{(q)}=\oplus_{p>q}V^p$. Then we have the $x$-invariant
filtration:
$$
V=V^{(0)}\supset V^{(1)}\supset\dots \supset V^{(\nu)}=0.
$$
Denote $V_{(q)}=\oplus_{p<q}V^p$. Then we have the
$x'$-invariant filtration:
$$
0=V_{(0)}\subset V_{(1)}\subset\dots \subset V_{(\nu)}=V.
$$
The vector space $V_{(q)}$ is a submodule in $(V,x')$.
Since $x'$ is $\Th_{V,\Om}$-stable we have $\Th_{V,\Om}(V_{(q)})>0$.
However $V_{(q)}$ is a quotient module in $(V,x)$.
Therefore, $(V,x)$ is unstable.
\end{proof}

\subsection{Stable construction}

\subsubsection{} 
Let us consider the cocycle $\e_{\Om}$ associated to our
fixed orientation $\Om$, see \cite{FrenkelKac1981}, \cite[1.3.4]{FMV00}.
By construction, the functions $E^{*}_{\al}$, ${\al}\in R_{+}$  
on $\Lambda_V$ defined as in \ref{LieEulerfinite} with the 
cocycle $\e=\e_{\Om}$
form a basis
of the Lie algebra $n^{*}$, and the $*$-bracket is given by:
\begin{equation}\label{F}
[E^{*}_\alpha , E^{*}_\beta ]=
\begin{cases} 
\epsilon (\alpha , \beta)
E^{*}_{\alpha + \beta} , 
&\text{ if } \alpha + \beta \in R_+ , \\
0 &\text{ if } \alpha + \beta \notin R_+ .
\end{cases}
\end{equation} 

\subsubsection{} It is clear that the affine variety 
$\E_{V,\Om}\subset \Lambda_V$ 
is one of the irreducible components of $\Lambda_V$.
We can define a constructible function $E_{\al}$, ${\al}\in R_{+}$  
on $\E_{V,\Om}$, $\dim V=\alpha$
as follows, see \cite[4.3.1.b]{FMV00}:
\begin{equation}\label{defEalpha}
E_{\al}=
\begin{cases} 
1, &\text{ if } x\in \E^{\Ind}_{V,\Om} , \\
0 &\text{ otherwise } .
\end{cases}
\end{equation}

According to \cite[4.3.4]{FMV00},
the space spanned by the functions $E_{\al}$, ${\al}\in R_{+}$
is a Lie algebra (which we will denote here by $\n^{\Om}$) 
with the bracket $[f,g ]_{\Om}=f*_{\Om}g-g*_{\Om}f$.
The map $\n^{\e}\to \n^{\Om}$
given by $\tilde e_{\al}\to E_{\al}$ is an isomorphism. In other words,
the bracket can
be explicitly described as follows:
\begin{equation}\label{E}
[E_\alpha , E_\beta ]_{\Om}=
\begin{cases} 
\epsilon (\alpha , \beta)
E_{\alpha + \beta} , 
&\text{ if } \alpha + \beta \in R_+ , \\
0 &\text{ if } \alpha + \beta \notin R_+ .
\end{cases}
\end{equation}

\subsubsection{} Let us take the character $\Th_{V,\Om}$, 
and denote the identity function on
$\Lambda_V^s$, $\dim V={\al}$, by 
${\widetilde E}_{\al}$, ${\widetilde E}_{\al}(\Lambda_V^s)\equiv 1$, 
where $\Lambda_V^s$ is the set of points in $\Lambda_V$ stable 
with respect to $\Th_{V,\Om}$. Notice that by Lemma \ref{stablemma},
$\Lambda_V^s=\E_{V,\Om}^{\Ind}$, and so the function ${\widetilde E}_{\al}$
is constant on a single $G_V$-orbit in $\Lambda_V^s$.

\begin{lemma}\label{restrictions} Up to a sign:
\begin{enumerate}
\item $E^{*}_{\al}|_{\E_{V,\Om}}=E_{\al}$,
\item $E^{*}_{\al}|_{\Lambda_V^s}={\widetilde E}_{\al}$.
\end{enumerate}
\end{lemma}

\begin{proof} Follows immediately from definitions,
the proof of \cite[Theorem 12.13]{Lusztig1991a},
and the formulas \ref{F}, \ref{E}.
\end{proof}

\subsubsection{} 
Let us denote the space spanned by the functions ${\widetilde E}_{\al}$, 
$\al\in R_{+}$ by $\n^{stable}$. Since $\Lambda_V^s=\E_{V,\Om}^{\Ind}$, 
by lemma \ref{stablemma},
the restrictions from $\Lambda_V$ to $\E_{V,\Om}$ 
to $\Lambda_V^s$ provide us with 
the based isomorphisms of vector spaces (cf. Lemma \ref{restrictions})
\[
\begin{CD}
\n^{*}@>{\simeq}>{\text{restriction}}>\n^{\Om}
@>{\simeq}>{\text{restriction}}>\n^{stable}, \\
\end{CD}
\]
defined by
$E^{*}_{\al} \mapsto E_{\al} \mapsto {\widetilde E}_{\al}$
for $\al\in R_{+}$. 
It is clear that the first restriction is an isomorphism of Lie 
algebras, and
the second restriction equips the space $\n^{stable}$ of
functions on 
$\Lambda_V^s$ constant on $G_V$-orbits (or equivalently,
functions on $\Lambda_V^s/G_V=\pt_V$)
with the structure of a Lie algebra isomorphic to 
$\n^{*}\simeq\n^{\Om}\simeq\n^{\e}$. 

\section{Stability, affine Lie algebras of type ADE, and a 
conjectural construction of the algebras of BPS states}

\subsection{Physics}

\subsubsection{} 
One of the current models of string theory defines $D$-branes
as objects in the derived category $D^b(Coh(Z))$ of the coherent sheaves
on the algebraic (Calabi-Yau) variety $Z$. Physicists 
consider the moduli space $\mathcal{M}_{\zeta}(n)$  (where 
$\zeta=(\zeta_1,\zeta_2,\dots,\zeta_N)$ are the Fayet-Iliopulos terms) of 
\emph{semistable} $D$-branes of charge $n$, cf. 
\cite{HM98, DFR00, FM00}. 
In the semiclassical approximation, 
the space of BPS states is given by the cohomology
of the moduli space:
\[
\mathcal{H}_{\text{BPS}}=H^*(\mathcal{M}_{\zeta}(n)).
\] 

\subsubsection{Dictionary}
We will consider the $D$-branes on the resolution 
$\widetilde{\C^2/\Gamma}$ of a simple
singularity $\C^2/\Gamma$. Let $Q$ be the extended Dynkin diagram 
corresponding to $\Gamma$ via the McKay correspondence.
We suggest a mathematical model of the physical situation in this case
with which we will work. In the next subsection \ref{justification} 
we will justify our model. For now, we provide the physics/mathematics
dictionary:

\emph{the charge lattice} : the positive part $\Z_{\geq 0}[I]$ of 
the affine root lattice associated to $Q$; 

\emph{places in the the charge lattice occupied by the 
single particle BPS states} : positive roots $R_{+}\subset\Z_{\geq 0}[I]$;

\emph{ a $D$-brane of charge $\alpha\in\Z[I]$ on 
$\widetilde{\C^2/\Gamma}$} : a representation 
$x\in\Lambda_V$, $\dim V=\alpha$ of the quiver (I,H)
(with the underlying non-oriented graph $Q$);

\emph{ Fayet-Iliopulos terms $\zeta$} : an additive function $c$ 
which defines Rudakov's slope stability together with 
$r(V)= \sum_{i\in I}\dim V_i$; 

\emph{ $\zeta$-stable $D$-branes} : stable elements of $\Lambda_V$ with
respect to a stability condition; 

\emph{ moduli spaces of $D$-branes of charge $\alpha\in R^{\re}_{+}$} :
$\Lambda^{s}_V/G_V=\pt_{\alpha}$, $\dim V=\alpha\in R^{\re}_{+}$; 

\emph{  moduli spaces of $D$-branes of charge 
$\delta=\text{indivisible imaginary root}$} :  
$\Lambda^{s}_V/G_V=\mathfrak{L}=\mathfrak{L}_1$, where $\mathfrak{L}$
is the exceptional fiber of the resolution of simple singularity;

\emph{ moduli spaces of $D$-branes of charge 
$m\delta$, $m>1$} : $\Lambda^{ssd}_V/(S-\text{equivalence})=
\mathfrak{L}=\mathfrak{L}_m$, where $\Lambda^{ssd}_V$ is the 
semistable diagonal (see \ref{diagonal} for the definition, 
cf. \cite {HM98}), and where $\mathfrak{L}$
is the exceptional fiber of the resolution of simple singularity;

\emph{ BPS states} : certain constructible functions on our 
moduli spaces, see \ref{bps} for details.

\subsubsection{Justification}\label{justification}
If we consider the $D$-branes on  
$\widetilde{\C^2/\Gamma}$ 
then due to M. Kapranov
and E. Vasserot \cite{KV}:
\[
D^b(Coh(\widetilde{\C^2/\Gamma}))\simeq D^b(Rep(Q))
\]
where $Q$ is the extended Dynkin diagram associated to
the $\Gamma\subset SL(2,\C)$ via the McKay correspondence,
and $Rep(Q)$ is the category of finite-dimensional 
\emph{double} representations of $Q$ in the sense of \cite[3.4]{KV}. 

We are only interested in $D$-branes of charge $\alpha\in R_{+}$ 
which are $0$-complexes on the 
right hand side, i.e. double representations of $Q$ rather than complexes
of such representations. Such representations are identified
with elements $x\in\E_{V}$, $\dim V=\alpha$ 
satisfying $m_i(x)=0$ for all $i\in I$
(see section \ref{moment} for the definition of $m_i$).
The notion of stability becomes the usual GIT stability adapted
to the quiver situation by King and Rudakov (see section 
\ref{stablesection}). We choose 
a ``non-degenerate'' stability condition
(cf. \ref{nondegenerate}).
 
If $\dim V=\alpha\in R^{\re}_{+}$, then the moduli space is a point:
$\mathcal{M}_{\zeta}(\alpha)=\pt_\alpha$.

If $\dim V=\delta$, where $\delta$ is the indivisible imaginary root,
then the moduli space is the resolution  of a simple singularity:
\[
\mathcal{M}_{\zeta}(\delta)=\widetilde{\C^2/\Gamma} .
\]
In the two cases above one may replace the (middle) cohomology
of the moduli spaces with the constructible functions on the 
exceptional fiber see \cite[10.16]{Nakajima1994}.
Following this logic, we suggest the interpretation of BPS states
as in the dictionary above and in \ref{bps} below  
if $\dim V=m\delta$, $m>1$ where the situation is more complicated 
and a clear connection between physical and mathematical results was not
available before, as far as we know.

\subsubsection{} Physics implies that the space of BPS states
on $\widetilde{\C^2/\Gamma}$
of all possible charges should form a Lie algebra isomorphic
to the plus part of the affine Lie algebra corresponding to $\Gamma$
via the McKay correspondence \cite{FM00}. 
The main purpose of this section is
to offer a conjectural, but mathematically rigorous, validation
of this claim in \ref{bps}.

\subsection{The conjecture}\label{generalcase}

In this section we consider quivers of affine ADE type, i.e. 
the underlying non-oriented graph $Q$ is an extended Dynkin graph
of ADE type.

\subsubsection{} Let us fix an extending vertex $p$ of the graph $Q$.
For every $\alpha=\dim V$ let us 
fix the Nakajima's character: $\theta(V)=-1$, if 
$\dim V=k\in I'=I-\{p\}$, and $\theta(V)=\sum_{i\in I'} \dim_{\C}(V_i)$
if $\dim V=p\in I$ (cf. \cite{Nakajima1998}).
(Semi)stable points in this section are considered with respect to 
this stability condition.  If $V_p=0$ we can take a King's character
$\Th_{V,\Om}$ (see \ref{king}) associated to any orientation that 
``flows to the extending vertex'' i.e. the extending vertex is a sink 
and one can get to the extending vertex from any other point in the quiver 
going along the oriented edges.
In the DE case there is only one
such orientation (once the extending vertex is fixed). 
In the $A_n$ case there are $n$ such orientations. 

\subsubsection{} For $\dim V=\alpha\in R_+^{\re}$
we have (cf. \cite{CB01}):
\begin{equation}
\Lambda^{s}_V/G_V=\pt_{\alpha} .
\end{equation}
Let us denote by ${\widetilde E}_{\alpha}$ the function identically $1$ 
on the point
$\pt_{\alpha}$, ${\widetilde E}_{\alpha}(\pt_{\alpha})\equiv 1$.

\subsubsection{} 
For $\dim V=\delta=\text{indivisible imaginary root}$ we have 
(\cite{Kronheimer1989}, 
\cite{Nakajima1994}):
\begin{equation}
\Lambda^{s}_V/G_V=\mathfrak{L} ,
\end{equation}
where $\mathfrak{L}=\mathfrak{L}_1$ is the exceptional fiber of 
the resolution of simple singularity 
$$
\mathfrak{L}\hookrightarrow \widetilde{\C^2/\Gamma}\to \C^2/\Gamma .
$$
Here $\Gamma$ is the finite subgroup of $SL(2,\C)$ corresponding
to the diagram $Q$ via the McKay correspondence. 

It is well known that $\mathfrak{L}$ is a configuration of lines 
$\mathbb{P}^1$ which are its irreducible components and which may be
indexed by the vertices $k\in I'=I-\{p\}$. We will denote the $k^{\text{th}}$
irreducible component of $\mathfrak{L}$ by $Y_k$, $k\in I'$.
Let us denote the characteristic function of $Y_k$ by ${\widetilde E}_k(1)$, 
$k\in I'$, ${\widetilde E}_k(1)(Y_k)\equiv 1$, ${\widetilde E}_k(1)(\mathfrak{L}-Y_k)\equiv 0$.

\subsubsection{} 
Finally, for $\dim V=m\delta$, $m>1$ we introduce the
\emph{semistable diagonal} $\Lambda^{ssd}_V\subset\Lambda_V$
as follows (cf. \cite{HM98}): $x$ is $S$-equivalent to the direct sum
of $m$ isomorphic representations in $\Lambda_{\delta}^s$. 
Here $\Lambda_{\delta}^s=\Lambda_{V'}^s$, $\dim V'=\delta$.
Formally: 
\begin{equation}\label{diagonal}
\Lambda^{ssd}_V=\{x\in \Lambda_V\ |\ x\overset{S}\sim 
x_1\oplus x_2\oplus\dots\oplus x_m, x_1\simeq x_2\simeq\dots\simeq x_m, 
x_i\in\Lambda_{\delta}^s \}
\end{equation}
It is clear that:
$$
\Lambda^{ssd}_V/(S\text{-equivalence})=\mathfrak{L} 
$$
where $\mathfrak{L}=\mathfrak{L}_m$ is the same exceptional variety as 
above.
Let us denote the characteristic function of the $k^{\text{th}}$ 
irreducible component $Y_k$ of $\mathfrak{L}=\mathfrak{L}_m$
by ${\widetilde E}_k(m)$, 
$k\in I'$, ${\widetilde E}_k(m)(Y_k)\equiv 1$, ${\widetilde E}_k(m)(\mathfrak{L}-Y_k)\equiv 0$.

\subsubsection{} 
We have defined the functions $E^{*}_{\alpha}$ on $\Lambda_V$, 
$\dim V=\al\in R^{\re}_{+}$, and the functions $E^{*}_k(m)$, 
$k\in I'=I-\{p\}$ on $\Lambda_V$, $\dim V=m\delta$, $m\geq 1$
in section \ref{LieEuleraffine}.
We need one more definition (cf. \ref{EpsilonCommutatorAffine}):
$$
\hat{E}^{*}_k(m):=\epsilon(\alpha_k,\beta_k)
[E^{*}_{\alpha_k}|_{\Lambda^{s}_{V_1}}, 
E^{*}_{\beta_k}|_{\Lambda^{s}_{V_2}}] ,
$$
where $\dim V_1={\alpha_k}=k\in I'\subset R_+^{\re}$, 
$\dim V_2={\beta_k}=(m\delta-\alpha_k)\in R_+^{\re}$
(here we treat elements of $I'$ as simple roots),
and the bracket is in the sense of
\ref{defl0}. 

Since the functions $E^{*}_{\alpha}, \hat{E}^{*}_k(m)$ are 
constant on $G_V$-orbits, we can consider them as 
functions on $\Lambda_V/G_V$. The equalities in the following
conjecture are understood in this sense.

\begin{conjecture}\label{conjecture} 
\begin{enumerate}
\item Up to a sign, 
$E^{*}_{\alpha}|_{\Lambda^{s}_V}={\widetilde E}_{\alpha}$, 
$\dim V={\alpha}\in R_+^{\re}$.
\item Up to a sign, 
$\hat{E}^{*}_k(1)|_{\Lambda^{s}_V}={\widetilde E}_k(1)$, 
$\dim V=\delta$.
\item Let $\dim V=m\delta$, $m>1$. 
We conjecture that if 
$x', x''\in \Lambda^{ssd}$
are two elements in the same $S$-equivalence class and
$\hat{E}^{*}_k(m)(x')\neq 0$, $\hat{E}^{*}_k(m)(x'')\neq 0$, 
then $\hat{E}^{*}_k(m)(x')=\hat{E}^{*}_k(m)(x'')$.
Thus we can consider $\hat{E}^{*}_k(m)|_{\Lambda^{ssd}_V}$ as a function on 
$\Lambda^{ssd}/S\text{-equivalence}$ by setting for an 
$S$-equivalence class $X$:
$$
\hat{E}^{*}_k(m)(X)=\begin{cases}
\hat{E}^{*}_k(m)(x), & \text{if there exists } x\in X \text{ with } 
\hat{E}^{*}_k(m)(x)\neq 0, \\
0, & \text{otherwise}.
\end{cases}
$$
\item Up to a sign, 
$\hat{E}^{*}_k(m)|_{\Lambda^{ssd}_V}={\widetilde E}_k(m)$, 
$\dim V=m\delta$.
\end{enumerate}
\end{conjecture}

The conjecture is verified in the $\hat A_1$ case.

\subsubsection{Remark}\label{nondegenerate} 
We would like to explain here that there is nothing special
about the Nakajima's choice of character.
If $\dim V=m\delta$ is an imaginary root,
then the space of all characters defining stability may be naturally
identified with a (classical) Cartan subalgebra 
$\mathfrak{h}=\C$-vector space spanned by ${i\in I'}$. 
A character $\tha$ is called non-degenerate if it does not
lie on a wall of a Weyl chamber.
We expect the conjecture to be true for any non-degenerate
character in the same Weyl chamber as the Nakajima's 
character (this is the fundamental chamber). 
Moreover, we could take a non-degenerate character in any other 
Weyl chamber,
but that would change our choice of simple roots which we identify with
the set $I'=I-\{p\}$. One could perhaps reformulate our conjecture in 
this case modifying the functions $E^{*}_k(m)$ using the action of 
the Weyl group.

\subsubsection{Remark} The functions ${\widetilde E}_k(m)$
parameterize the set $\Irr\mathfrak{L}=\Irr\mathfrak{L}_m$   
of the irreducible components of the exceptional variety $\mathfrak{L}$.
The components of $\mathfrak{L}$ are a basis in the space 
$H^2(\widetilde{\C^2/\Gamma})$, see \cite{Nakajima1994}.

\subsubsection{}\label{bps} 
Denote the Lie algebra spanned
by the functions $E^{*}_{\alpha}$ on  $\Lambda_V$, 
$\dim V=\al\in R^{\re}_{+}$, and the functions $E^{*}_k(m)$
$k\in I'=I-\{p\}$ on $\Lambda_V$, $\dim V=m\delta$, $m\geq 1$
by $n^{*}(Q)$, cf. section \ref{LieEuleraffine}.

Let us denote the vector space spanned by
the functions ${\widetilde E}_{\al}$, $\al\in R_{+}^{\re}$, and the functions 
${\widetilde E}_k(m)$ on 
$\mathfrak{L}=\mathfrak{L}_m=\Lambda_{m\delta}^{ssd}/(S-\text{equivalence})$
by $\n^{stable}(Q)$. This is the vector space of constructible functions on 
moduli spaces of (semi)stable points in $\Lambda$, which are linear 
combinations of characteristic functions of irreducible components.
The space $\n^{stable}(Q)$ is our model of the space of BPS states
at the orbifold $\C^{2}/\Gamma$, with $Q$ corresponding to $\Gamma$ 
via the McKay correspondence.

The conjecture, if true, would imply that the restriction 
from $\Lambda$ to $\Lambda^{s}$ (or $\Lambda^{ssd}$) gives us the 
based isomorphism of vector spaces
\[
\begin{CD}
\n^{*}(Q)@>{\simeq}>>\n^{stable}(Q) 
\end{CD}
\]
such that 
$E^{*}_{\al} \mapsto {\widetilde E}_{\al}$ and
$E^{*}_k(m)  \mapsto {\hat E}_k(m) \mapsto {\widetilde E}_k(m)$
for $\al\in R_{+}^{\re}$, $k\in I'$, $m\geq 1$. 
This map equips
the space of BPS states $\n^{stable}(Q)$ with the structure of a Lie algebra
isomorphic to the plus part of the affine Lie algebra corresponding 
to $Q$.

On the other hand the above isomorphism provides 
the plus part of the affine Lie algebra corresponding to $Q$,
with a basis $\{{\widetilde E}_{\al},\al\in R_{+}^{\re}; {\widetilde E}_k(m),k\in I',m\in\Z_{>0}\}$ 
indexed by irreducible components of algebraic varieties 
$\pt_{\al}=\Lambda_{\al}^{s}/G_{\al}$, and 
$\mathfrak{L}_m=\Lambda_{m\delta}^{ssd}/(S-\text{equivalence})$, 
$m\in\Z_{>0}$.

\section{Remarks on semicanonical basis for simple Lie algebras}
\label{semicanonical}

\subsection{Semicanonical basis}\label{semicanonicalbasis}
This subsection is lifted from \cite[2.4-5]{Lusztig1999}.
Recall the setup of \ref{georealization}. Let 
$V=\oplus_{i\in I}V_i$ be a $I$-graded vector space.
If $Y\in\Irr\Lambda_V$ is an irreducible component of $\Lambda_V$
and $f\in\MLZ(V)$ then there is a unique $c\in\Z$ such that
$f^{-1}(c)\cap Y$ contains an open dense subset of $Y$. 
Note that $f\mapsto c$ is a linear function $\rho_Y:\MLZ(V)\to\Z$. 
  
\begin{lemma} Let $Y\in\Irr\Lambda_V$.
There exists $f=e_Y\in\psi_V(U^{+}_{\Z,V})$ such that $\rho_Y(f)=1$
and $\rho_{Y'}(f)=0$ for any $Y'\in\Irr\Lambda_V-\{Y\}$.
\end{lemma}

Lusztig proves that the functions $e_Y$, $Y\in\Irr\Lambda_V$ 
form a basis in $\psi_V(U^{+}_{V})$, and therefore, the collection
of $e_Y$, $Y\in\Irr\Lambda_V$ for all possible dimensions of 
$V$ forms a basis in
the algebra $\psi(U^{+})=\ML_0$ isomorphic to $U^{+}$. This basis
is called \emph{semicanonical}.

In the remainder of this section we consider quivers of finite ADE type.

\subsubsection{}\label{rootdecompositions}
Since our quiver is of finite ADE type we know \cite[14.2]{Lusztig1991a}
that the irreducible components of $\Lambda_V$ are the closures
of conormal bundles of the various $G_V$-orbits in $\E_{V,\Om}$,
where $\Om$ is some orientation. Such $G_V$-orbits are indexed
by the decomposition of $(V,x)$, $x\in\E_{V,\Om}$ into a direct sum
of indecomposable submodules $(V,x)=(V_1,x_1)\oplus\dots\oplus(V_l,x_l)$
with $x_i\in\E^{\Ind}_{V_i,\Om}$. (The decompositions are considered up to
the order of the summands.) Thus, there is a one-to one correspondence 
between $\Irr\Lambda_V$ and the decompositions 
$\dim V=\dim V_1+\dots +\dim V_l$ of $\dim V\in\Z[I]$
into a sum of positive roots $\dim V_i\in R_{+}$. 
(Recall that $\E^{\Ind}_{V,\Om}\neq\emptyset$ if and only if 
$\dim V\in R_{+}$.) Thus the elements of the semicanonical basis
may be indexed by decompositions of $\dim V$ into a sum of positive roots.

\subsection{Open problem}\label{openproblem} 
Recall that we have defined the functions 
$E^{*}_{\al}$ on $\Lambda_V$, $\dim V=\al\in R_+$ in 
\ref{LieEulerfinite}. By construction
these functions are defined canonically up to a sign which
depends on the choice of the cocycle $\e$. Since 
$E^{*}_{\al}\in\psi_{V}(U^{+}_{\Z,\dim V})$, we can decompose 
$E^{*}_{\al}$
with respect to the semicanonical basis:
\begin{equation}\label{semidecomposition}
E^{*}_{\alpha}=\sum_{Y\in\Irr\Lambda_V} c_Y e_Y .
\end{equation}
The problem is to calculate the integer coefficients $c_{Y}$
for all irreducible components $Y$ of $\Lambda_V$. These 
coefficients are independent of any choices up to simultaneous 
multiplication of all of them by $-1$. Suppose that
an irreducible component $Y\in\Irr\Lambda_V$, $\dim V=\al$ 
corresponds to a decomposition 
\begin{equation}\label{alphapartition}
\al=\al_1+\dots+\al_l
\end{equation}
of $\al$ into
a sum of positive roots $\al_1,\dots,\al_l$ . 
Then the formula \ref{semidecomposition} may be 
regarded as an assignment 
$\{\al=\al_1+\dots+\al_l\}\mapsto c_Y$ of the integer $c_Y$ 
to every decomposition \ref{alphapartition}. It would be very interesting
to describe this assignment in terms of combinatorics of the root system 
without appealing to the semicanonical basis. We obtain such a description
in the $A_n$ case using the Weyl group $W=S_{n+1}$.
We also succeed in calculating $c_{Y}$ when $Y=\E_{V,\Om}$ are irreducible
components ``arising from orientations'', and we calculate the examples 
of $D_4$ and $D_5$.  

\subsection{Irreducible components arising from orientations}

\subsubsection{} Let us define the subvariety $O_V\subseteq\Lambda_V$
as follows:
\[
O_V=\bigcup_{\Om} \E_{V,\Om}
\]
where the union is taken over all orientations $\Om$ of our quiver.
Since each $Y=\E_{V,\Om}$ is an irreducible component of $\Lambda_V$
(we will call such $Y$ irreducible components arising from orientations),
we have $\Irr O_V\subseteq \Irr\Lambda_V$.
We will split the sum \ref{semidecomposition} into two parts:
\[
E^{*}_{\alpha}=E'_{\alpha}+E''_{\alpha}
\]
where
\begin{equation}
E'_{\alpha} =  \sum_{Y\in\Irr O_V} c_Y e_Y \qquad\text{and}\qquad
E''_{\alpha} =  \sum_{Y\in\Irr\Lambda_V-\Irr O_V} c_Y e_Y
\end{equation}

\subsubsection{}
If a root $\alpha\in R_{+}$ is not simple, represent it as
a sum of simple roots $\alpha_{k_1},\alpha_{k_2},\dots ,\alpha_{k_h}$,
$k_l\in I$, where $h=\height\alpha$
\begin{equation}\label{alphapresentation}
\alpha=\alpha_{k_1}+\alpha_{k_2}+\dots + \alpha_{k_h}
\end{equation}
in such a way that $\alpha_{k_1}+\alpha_{k_2}+\dots + \alpha_{k_j}$
is a root for $1\leq j\leq h$. 

It follows from our construction and definitions that up to a sign 
\begin{equation}\label{Ealpha}
E^{*}_{\alpha}=
[\dots [E^{*}_{\alpha_{k_1}},E^{*}_{\alpha_{k_2}}]\dots E^{*}_{\alpha_{k_h}}]
\end{equation}

Now we can calculate the coefficients of $E'_{\alpha}$
(cf. section \ref{stablesection} for a similar discussion).

\begin{proposition}\label{irrorientations} 
Up to a sign, 
\begin{equation}\nonumber
E'_{\alpha}=\sum_{Y=\E_{V,\Om}\in\Irr O_V} 
(\prod_{i<j}\epsilon_{\Om}(\alpha_{k_i}, \alpha_{k_j})) e_Y , 
\end{equation}
where $\epsilon_{\Om}$ is the Frenkel-Kac cocycle corresponding 
to $\Om$, \cite[1.3.4]{FMV00}.
This presentation does not depend on the
decomposition \ref{alphapresentation}.
\end{proposition}

\begin{proof} 
Let us fix some orientation $\Om$,
and let us assume that $Y=\E_{V,\Om}$. 
According to Lusztig \cite[12.13]{Lusztig1992}
we have a homomorphism of algebras $\ML_0\to\ML_{0,\Om}$
induced by the restriction on functions from $\Lambda_V$
to $\E_{V,\Om}$ (see \ref{defl0} for the definitions of 
$\ML_0$ and $\ML_{0,\Om}$). Then
\begin{equation}\label{allorientations}
E^{*}_{\alpha}|_Y =E'_{\alpha}|_Y=
[\dots [E_{k_1},E_{k_2}]_{\Om}\dots E_{k_h}]_{\Om}
=\prod_{i<j}\epsilon_{\Om}(\alpha_{k_i}, \alpha_{k_j})E_{\alpha},
\end{equation}
where $E_{\alpha}$ is defined by \ref{defEalpha},
and the last equality follows from \cite[4.3.4.b]{FMV00}.
Now it is clear that $\rho_Y(E^{*}_{\alpha})=\rho_Y(E^{*}_{\alpha}|_Y)$.
Then
\begin{equation}\nonumber
c_Y=\rho_Y(E^{*}_{\alpha})=\rho_Y(E^{*}_{\alpha}|_Y)= 
\prod_{i<j}\epsilon_{\Om}(\alpha_{k_i}, \alpha_{k_j})
\rho_Y(E_{\alpha,\Om}) 
= \prod_{i<j}\epsilon_{\Om}(\alpha_{k_i}, \alpha_{k_j}) .
\end{equation}
Notice that due to the formula \ref{allorientations} we 
know explicitly $E^{*}_{\alpha}|_O$ i.e., 
the value $E^{*}_{\alpha}(x)$ at any point $x\in O\subseteq\Lambda$.
\end{proof}

\subsubsection{}
Let $Q$ be the Dynkin graph
of type $A_n$.  
In this case it is clear that for $V\in\mathcal V$, 
$\dim V=\alpha\in R_{+}$
any irreducible component of $\Lambda_V$ is of the form $\E_{V,\Om}$
for some orientation $\Om$. 
Thus Proposition \ref{irrorientations} gives a complete answer 
to our question \ref{openproblem} in this case.  

\subsection{Conjugacy classes of the Weyl group}

\subsubsection{}\label{componentstoclasses} 
Recall the isomorphism 
$\psi: \n\to \n^*$ (see \ref{psi}), and let us take 
${\tilde e}_{\alpha}\in\n^{\e}\simeq\n$ 
such that $\psi({\tilde e}_{\alpha})=E^{*}_{\alpha}$,
$\alpha\in R_+$, 
as in \ref{LieEulerfinite}.
Let us fix $Y\in\Irr\Lambda_V$ and let 
$\alpha=\beta_1+\dots+\beta_l$ be the corresponding decomposition 
of $\alpha$ 
into the sum of positive roots.
Now consider the element 
${\tilde e}_{\beta_1}+\dots+{\tilde e}_{\beta_l}\in\n^{\e}\simeq\n$, 
and consider its
adjoint orbit. In such a way we obtain a map from $\Irr\Lambda_V$
to the set $\mathcal N$ of nilpotent orbits of the Lie algebra
corresponding to the Dynkin diagram $Q$. Moreover, following
Kazhdan-Lusztig \cite{KL} and Spaltenstein \cite{Spal},
we can construct a map from $\mathcal N$
to the set $\Cl (W)$ of conjugacy classes of the Weyl group $W$.
Thus we obtain a map $\cl :\Irr\Lambda_V\to \Cl (W)$ for any 
$V\in\mathcal V$.
We will use this map in the following subsection.

\subsubsection{Example: $A_n$} 

\begin{proposition}\label{signcharacter}
Up to a sign,
$$
E^{*}_{\alpha}=\sum_{Y\in\Irr\Lambda_V} \sgn(\cl Y)\  e_Y\ ,
$$
where $\sgn(\cl Y)$ is the value of the one-dimensional sign
character of $W=S_{n+1}$ on the conjugacy class $\cl Y$.
\end{proposition}

\begin{proof} The proof is straightforward and is left to the reader.
\end{proof}

\subsubsection{Open question} We have the map $\cl: \Irr\Lambda_V\to\Cl(W)$
for all ADE quivers, see \ref{componentstoclasses}. 
Looking at the $A_n$ case 
(Proposition \ref{signcharacter}), one may ask
if in the DE case the coefficients $c_{Y}$ are also governed by 
characters of the Weyl group. Unfortunately, the authors' naive attempt 
to replicate the $A_n$ result fails already in the $D_4$ case
(however the explicit calculation for the maximal root of $D_4$ is given 
below).

\subsection{Example: $D_4$} 

\subsubsection{} We index the vertices of the Dynkin diagram as follows:

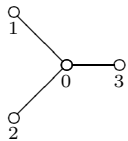
\begin{figure}[ht]
\begin{align}\nonumber
\xygraph{
[] 
!{<0pt,0pt>;<20pt,0pt>:} 
[u] 
*\cir<2pt>{}  
!{\save -<0pt,6pt>*\txt{$_1$}  \restore}
- [rd] 
*\cir<2pt>{}
!{\save -<0pt,6pt>*\txt{$_0$}  \restore}
- [ld] 
*\cir<2pt>{} 
!{\save -<0pt,6pt>*\txt{$_2$}  \restore}
[ru] 
*\cir<2pt>{}
- [r] 
*\cir<2pt>{}
!{\save -<0pt,6pt>*\txt{$_3$}  \restore}
%- [r] 
%*\cir<2pt>{}
%!{\save -<0pt,6pt>*\txt{$_4$}  \restore}
}
\end{align}
\end{figure}
The simple roots are denoted by 
$\alpha_0, \alpha_1, \alpha_2, \alpha_3$.
Let $\alpha  =2\alpha_0+\alpha_1+\alpha_2+\alpha_3$ 
be the maximal root. Indexing the irreducible components
of $\Lambda_V$, $\dim V=\alpha$ by the decompositions of $\alpha$ 
we obtain by a straighforward computation using
the properties of the semicanonical basis:
$$
\begin{aligned}
E^{*}_{\alpha}& = E'_{\alpha}+E''_{\alpha} \\
&= E'_{\alpha}- e_{(\alpha_0+\alpha_1)+(\alpha_0+\alpha_2+\alpha_3)}
- e_{(\alpha_0+\alpha_2)+(\alpha_0+\alpha_1+\alpha_3)}
- e_{(\alpha_0+\alpha_3)+(\alpha_0+\alpha_1+\alpha_2)} \\
& - e_{\alpha_0+\alpha_1+(\alpha_0+\alpha_2+\alpha_3)}
- e_{\alpha_0+\alpha_2+(\alpha_0+\alpha_1+\alpha_3)}
- e_{\alpha_0+\alpha_3+(\alpha_0+\alpha_1+\alpha_2)} \\
& +2 e_{\alpha_0+(\alpha_0+\alpha_1+\alpha_2+\alpha_3)}.
\end{aligned}
$$

\subsection{Example: $D_5$} 

\subsubsection{} We index the vertices of the Dynkin diagram as follows:
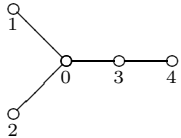
\begin{figure}[ht]
\begin{align}\nonumber
\xygraph{
[] 
!{<0pt,0pt>;<20pt,0pt>:} 
[u] 
*\cir<2pt>{}  
!{\save -<0pt,6pt>*\txt{$_1$}  \restore}
- [rd] 
*\cir<2pt>{}
!{\save -<0pt,6pt>*\txt{$_0$}  \restore}
- [ld] 
*\cir<2pt>{} 
!{\save -<0pt,6pt>*\txt{$_2$}  \restore}
[ru] 
*\cir<2pt>{}
- [r] 
*\cir<2pt>{}
!{\save -<0pt,6pt>*\txt{$_3$}  \restore}
- [r] 
*\cir<2pt>{}
!{\save -<0pt,6pt>*\txt{$_4$}  \restore}
}
\end{align}
\end{figure}

The simple roots are denoted by 
$\alpha_0, \alpha_1, \alpha_2, \alpha_3, \alpha_4$.
Let $\alpha=2\alpha_0+\alpha_1+\alpha_2+2\alpha_3+\alpha_4$
be the maximal root.
There are $55$ root partitions of $\alpha$, e.g.
$\alpha=\al_0+\al_3+(\al_0+\al_1+\al_2+\al_3+\al_4)$.
Indexing the irreducible components
of $\Lambda_V$, $\dim V=\alpha$ by the decompositions of $\alpha$ 
we obtain by a tedious but straightforward computation using
the properties of the semicanonical basis:
$$
\begin{matrix}
E^{*}_{\alpha} & = & E'_{\al}+E''_{\al}=E'_{\al}+ \\
&+&
e_{(2\al_0+\al_1+\al_2+\al_3)+(\al_3+\al_4)} 
&+&
e_{\al_2+(\al_0+\al_1)+(\al_0+\al_3)+(\al_3+\al_4)} \\
&+&
e_{\al_1+(\al_0+\al_2)+(\al_0+\al_3)+(\al_3+\al_4)}
&+&
e_{\al_1+\al_2+(\al_0+\al_3)+\al_0+(\al_3+\al_4)} \\
&+&
e_{(2\al_0+\al_1+\al_2+\al_3)+\al_3+\al_4} 
&+&
e_{\al_2+(\al_0+\al_1)+\al_3+\al_4+(\al_0+\al_3)} \\
&+&
e_{\al_1+(\al_0+\al_2)+(\al_0+\al_3)+\al_3+\al_4} 
&+&
e_{(\al_0+\al_3)+\al_0+\al_2+\al_3+\al_4} \\
&+&
e_{(\al_2+\al_0+\al_3)+(\al_1+\al_0+\al_3+\al_4)} 
 &+&
e_{(\al_1+\al_2+\al_0+\al_3)+(\al_0+\al_3)+\al_4} \\
&-&
e_{(\al_2+\al_0+\al_1)+(\al_0+\al_3)+(\al_3+\al_4)}
&+& 
e_{(\al_2+\al_0+\al_1)+\al_0+\al_3+(\al_3+\al_4)} \\
&+&
e_{(\al_2+\al_0+\al_1)+\al_0+2\al_3+\al_4} 
&-&
e_{(\al_0+\al_1+\al_2)+(\al_0+\al_3)+\al_4} \\
&-&
e_{(\al_0+\al_2)+(\al_0+\al_1+\al_3)+(\al_3+\al_4)} 
&+&
e_{(\al_0+\al_1+\al_3)+(\al_0+\al_2+\al_3+\al_4)} \\
&-& 
e_{\al_2+(\al_0+\al_1+\al_3)+\al_0+(\al_3+\al_4)} 
&+&
e_{\al_2+(\al_0+\al_1+\al_3)+(\al_0+\al_3+\al_4)} \\
&-&
e_{(\al_0+\al_1)+(\al_2+\al_0+\al_3)+(\al_3+\al_4)} 
&+&
e_{(\al_2+\al_0+\al_3)+(\al_0+\al_1+\al_3+\al_4)} \\
&-&
e_{\al_0+\al_1+(\al_2+\al_0+\al_3)+(\al_3+\al_4)} 
&+&
e_{\al_1+(\al_2+\al_0+\al_3)+(\al_0+\al_3+\al_4)} \\
&+&
e_{(\al_0+\al_2)+\al_3+(\al_0+\al_1+\al_3+\al_4)}
&-&
e_{(\al_0+\al_2)+\al_3+(\al_0+\al_1+\al_3)+\al_4} \\
&+&
e_{\al_0+\al_2+\al_3+(\al_0+\al_1+\al_3+\al_4)}
&-&
e_{\al_0+\al_2+\al_3+(\al_0+\al_1+\al_3)+\al_4} \\
&+&
e_{\al_3+(\al_0+\al_1)+(\al_0+\al_2+\al_3+\al_4)} 
&-&
e_{\al_3+(\al_0+\al_1)+(\al_0+\al_2+\al_3)+\al_4} \\
&+&
e_{\al_3+\al_0+\al_1+(\al_0+\al_2+\al_3+\al_4)} 
&-&
e_{\al_3+\al_0+\al_1+(\al_0+\al_2+\al_3)+\al_4} \\
&+&
2
e_{\al_0+(\al_3+\al_4)+(\al_0+\al_1+\al_2+\al_3)} 
&-&
2
e_{(\al_0+\al_3+\al_4)+(\al_0+\al_1+\al_2+\al_3)} \\ 
&-&
2
e_{\al_0+\al_3+(\al_0+\al_1+\al_2+\al_3+\al_4)}
&+&
2
e_{\al_0+\al_3+\al_4+(\al_0+\al_1+\al_2+\al_3)}\\
&-&
2
e_{(2\al_0+\al_1+\al_2\al_3+\al_4)+\al_3)}
&-&
2
e_{\al_2+(\al_0+\al_1)+(\al_0+\al_3+\al_4)+\al_3} \\
&-&
2
e_{\al_1+(\al_0+\al_2)+(\al_0+\al_3+\al_4)+\al_3}
&-&
2
e_{(\al_0+\al_3+\al_4)+\al_0+\al_1+\al_2+\al_3} \\ 
&+&
2 
e_{(\al_0+\al_3+\al_4)+(\al_0+\al_1+\al_2)+\al_3} &&  
\end{matrix}
$$

\subsubsection{} While in the $D_5$ case the coefficients in the
decomposition for the maximal root are $\pm 1, \pm 2$, in the $D_6$ case
the coefficients are  $\pm 1, \pm 2, \pm 4$. We expect
that in the $D_n$ case all the coefficients are the powers of $2$.

\subsection{More open questions}
The questions of decomposing ``canonically'' defined basis elements
of affine Lie algebra (such as $E^{*}_{\al}, E^{*}_k(n)$, see 
\ref{LieEuleraffine}) 
with respect to semicanonical basis make sense for any 
affine Lie algebra of type ADE.
It would be very interesting to develop a purely combinatorial
approach to these questions in the affine case as well.

%\bibliographystyle{amsalpha}
%\bibliography{hall}

\providecommand{\bysame}{\leavevmode\hbox to3em{\hrulefill}\thinspace}

\end{document}